\documentclass[conference]{IEEEtran}
\IEEEoverridecommandlockouts
\usepackage{cite}
\usepackage{amsmath,amssymb,amsfonts}
\usepackage{graphicx}
\usepackage{textcomp}
\usepackage{xcolor}

\usepackage{algorithm}
\usepackage{algcompatible}
\usepackage{algpseudocode}

\usepackage{tabularx}
\usepackage{tabularray}
\usepackage{setspace}
\usepackage{makecell}
\usepackage{multirow}
\usepackage{booktabs}
\usepackage{subcaption}
\usepackage{ifpdf}
\usepackage[cmintegrals]{newtxmath}
\usepackage{bm}
\usepackage{comment}
\def\BibTeX{{\rm B\kern-.05em{\sc i\kern-.025em b}\kern-.08em
    T\kern-.1667em\lower.7ex\hbox{E}\kern-.125emX}}
\newtheorem{definition}{Definition} 
\begin{document}

\title{KEWS: A KPIs-Based Evaluation Framework of Workload Simulation On Microservice System\\
}
\author{\IEEEauthorblockN{Pengsheng Li, Qingfeng Du$^\ast$\thanks{$^\ast$Corresponding author: Qingfeng Du}, and~Shengjie Zhao}
\IEEEauthorblockA{\textit{School of Software Engineering, Tongji University} \\
Shanghai, China \\
\{2131514, du\_cloud, shengjiezhao\}@tongji.edu.cn}
}

\maketitle

\begin{abstract}

Simulating the workload is an essential procedure in microservice systems as it helps augment realistic workloads whilst safeguarding user privacy. The efficacy of such simulation depends on its dynamic assessment. The straightforward and most efficient approach to this is comparing the original workload with the simulated one using Key Performance Indicators (KPIs), which capture the state of the system. Nonetheless, due to the extensive volume and complexity of KPIs, fully evaluating them is not feasible, and measuring their similarity poses a significant challenge. This paper introduces a similarity metric algorithm for KPIs, the Extended Shape-Based Distance (ESBD), which gauges similarity in both shape and intensity. Additionally, we propose a KPI-based Evaluation Framework for Workload Simulations (KEWS), comprising three modules: preprocessing, compression, and evaluation. These methodologies effectively counteract the adverse effects of KPIs' characteristics and offer a holistic evaluation. Experimental results substantiate the effectiveness of both ESBD and KEWS.


\end{abstract}

\begin{IEEEkeywords}
workload simulation, key performance indicators, microservice system, similarity metric
\end{IEEEkeywords}
\section{Introduction}

Workload simulation serves as a vital task to augment the realistic workload, the fundamental data source for AI operations (AIOps) in microservice systems~\cite{bao2023aiops}. Such workloads are crucial for various applications such as anomaly detection and root cause analysis~\cite{notaro2021survey}. This task involves the extraction and processing of features from the realistic workload, enabling the generation of a simulated workload bearing similar properties. Evaluating workload simulations necessitates not just static comparisons of workload characteristics~\cite{vogele2018wessbas}, but also dynamic assessments of system status under both realistic and simulated workloads. Key Performance Indicators (KPIs), which record the internal status of systems in a structured time-series format~\cite{meng2020localizing}, are indisputably the most typical monitoring data within microservice systems~\cite{lee2023eadro}. Nevertheless, three significant challenges confront the evaluation of workload simulations using KPIs in microservice systems.

Firstly, monitoring an enormous quantity of KPIs within microservice systems for real-time anomaly detection renders their use impractical for evaluation purposes. Secondly, the characteristics of KPIs are more complex relative to ideal time series, entailing various elements including amplitude differences, phase shifts, inevitable noise, high dimensionality, and large numerical spans. These characteristics alter the shape and intensity of the KPIs, thereby warping the similarities. Thirdly, the intricate characteristics of KPIs result in a shortage of suitable similarity metrics applicable to them.

To address these issues, we introduce an extended shape-based distance (ESBD) algorithm as a similarity metric for KPIs, taking into account both shape and intensity. Moreover, we examine a novel framework for workload simulation evaluation, which is the \textbf{K}PIs-based \textbf{E}valuation Framework of \textbf{W}orkload \textbf{S}imulation (KEWS). Our approach begins with preprocessing KPIs to diminish their complex characteristics. Subsequently, we generate representative sets of KPIs using a Domain Knowledge Filter (DKF), a Chaos Experiment Filter (CEF), and a Dynamic Time Warping (DTW)-based Density Adaptive DBSCAN Cluster. Finally, with these KPI sets, we assess the quality of the workload simulation employing ESBD. Our contributions can be summarized as follows:
\begin{itemize}
    \item We introduce a novel similarity metric algorithm for KPIs, ESBD, which discerns similarity by evaluating both shape and intensity.
    \item We propose a new, more practical framework for assessing workload simulation than currently existing models. To the best of our knowledge, this stands as the first attempt to evaluate workload simulation using KPIs.
    \item We conduct comprehensive experiments to corroborate the effectiveness of both ESBD and KEWS by comparing these with baseline methods.
\end{itemize}

\section{Related Work}

\subsection{KPI Similarity Metric}

The KPI similarity metric is widely used as both an absolute criterion for making statistical inferences regarding KPI interrelationships and a relative measure for downstream tasks\cite{lhermitte2011comparison, serra2014empirical}, which can be categorized primarily into two types: raw-based and feature-based methods. Raw-based methods directly calculate the similarity of KPIs, drawing on approaches such as Shape-based Distance (SBD)\cite{paparrizos2015k} and Dynamic Time Warping (DTW)\cite{berndt1994using, lines2015time}. Alternatively, feature-based methods map KPIs into low-dimensional latent spaces to learn representations for similarity, deploying techniques like the transformer-based method\cite{zerveas2021transformer} and contrastive learning method \cite{yue2022ts2vec}. Even though raw-based methods are better suited for precise workload evaluation, a majority of current techniques focus only on shape similarity while neglecting intensity.

\subsection{Workload Simulation Evaluation}

Being among the minority of studies addressing workload simulation evaluation, WESSBAS\cite{vogele2018wessbas}  evaluates workload using static statistics, thereby overlooking dynamic features. A crucial aspect of evaluating workload simulation dynamically lies in the discerning selection of suitable KPIs from a vast pool\cite{peiris2014pad}. Given the explicit or subtle correlations between KPIs\cite{meng2020localizing}, one can extract homogeneous KPIs through clustering. In recent times, there's been a fresh line of research investigating clustering algorithms for KPIs, such as the K-shape based method\cite{qian2020large}, hierarchical-based method\cite{halawa2020unsupervised}, and density-based methods\cite{li2018robust,wang2021rapid,yu2019unsupervised}. However, these methods focus solely on the value characteristics and disregard the intrinsic attribute characteristics of the KPI, which results in an information loss.
\section{Problem definition}
In this section, we present a set of formal descriptions for quality evaluation of workload simulation.

\begin{definition}[\textbf{Workload Simulation}]
    Given a workload $\mathcal{W}$ of a microservice system $\mathcal{M}$, workload simulation aims to generate a new workload $\mathcal{W}^{'}$ of $\mathcal{M}$ with properties of $\mathcal{W}$ .
\end{definition}

\begin{definition}[\textbf{Workload Evaluation}]
    Given data $\mathcal{D}, \mathcal{D}^{'}$ obtained by injecting original $\mathcal{W}$ and simulated workload $\mathcal{W}^{'}$ into the system $\mathcal{M}$, the target of workload evaluation $h(\mathcal{D}, \mathcal{D}^{'})$ is to measure the workload similarity between $\mathcal{W}$ and $\mathcal{W}^{'}$.
\end{definition}

\begin{definition}[\textbf{Similarity Metric}]
    Given two time series $\bm{x}\in\mathbb{R}^{m_x}, \bm{y}\in\mathbb{R}^{m_y}$,  the target of similarity metric $\phi(\bm{x}, \bm{y})$ is to calculate the similarity $\Phi$ sparing the shape variations, e.g., noise, amplitude differences and phase shifts.
\end{definition}
\section{Methodology}

\begin{figure*}[htbp]
    \centering
    \includegraphics[width=0.8\textwidth]{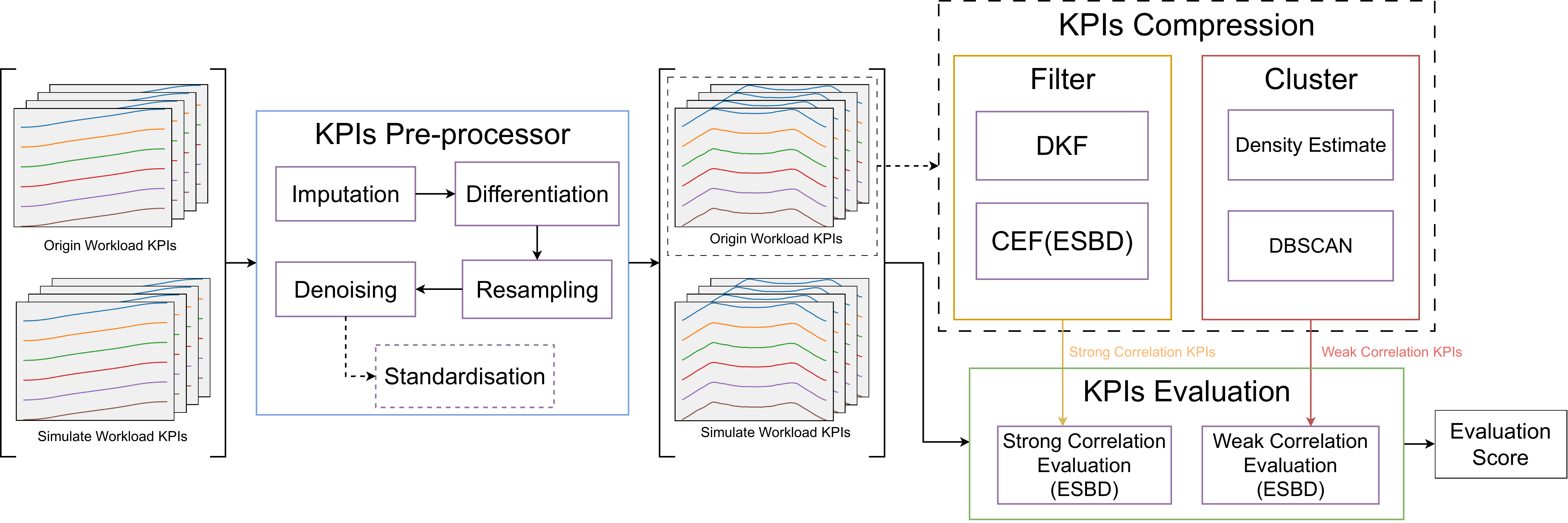}
    \caption{The overall pipeline of KEWS.}
    \label{fig:kews_overview}
    \vspace{-5mm}
\end{figure*}

As illustrated in Fig.\ref{fig:kews_overview}, KEWS comprises three components: pre-processing, compression, and evaluation. Initially, the corresponding KPIs are produced by injecting both the original workload $\mathcal{W}$ and the simulated workload $\mathcal{W}^{'}$ into the system. Following this, the KPIs are refined into increment-centric, low-noise, standardized data during the pre-processing stage. Subsequently, the compression of KPIs leverages original workload KPIs to condense the KPIs evaluation set. This process utilizes domain knowledge and chaos experiments in conjunction with ESBD for strong correlation and employs a DTW-based Density Adaptive DBSCAN clustering model for weak correlation. Finally, in the KPIs evaluation phase, ESBD is utilized to evaluate the similarity between the original and the simulated workload. The results are then aggregated to characterize the overall quality of the workload simulation.

\subsection{Similarity Metric}\label{Similarity Metric}
We propose the Extended-Shape-Based Distance (ESBD) as a comprehensive similarity metric for KPIs across varying scales. This consistent metric concurrently measures similarity in both shape and intensity. For ease of understanding, we exemplify ESBD using scenarios where the lengths of the KPIs are identical. It is crucial to note that ESBD is equally applicable in situations where these lengths differ. Formally, let $\bm{x}_1,\bm{x}_2\in\mathbb{R}^m$ denote two KPI time series of length $m$, as:
\vspace{-1mm}
\begin{equation}
    \phi _{\textit{esbd}}(\bm{x}_1, \bm{x}_2):= (1-\alpha)\phi _{\textit{shape}}(\bm{x}_1, \bm{x}_2) + 2\alpha \phi _{\textit{intensity}}(\bm{x}_1, \bm{x}_2),\label{eq:esbd}
    \vspace{-1mm}
\end{equation}
where $\phi _{\textit{shape}}(\cdot)$ measures the shape similarity of the KPIs, while $\phi _{\textit{intensity}}(\cdot)$ gauges their intensity similarity.We assign $\alpha\in[0, 1]$ as the intensity factor, which controls the balance between shape and intensity. Importantly, $\phi _{\textit{shape}}(\cdot)$ employs the cross-correlation-based SBD method\cite{paparrizos2015k} to depict KPI shape similarity, which can be formulated as:
\vspace{-1mm}
\begin{equation}
    \phi_{\textit{shape}}(\bm{x}_1, \bm{x}_2):=1-\max _{l}\left(\frac{R_{l-m}(\bm{x}_1, \bm{x}_2)}{\sqrt{R_{0}(\bm{x}_1, \bm{x}_1) \cdot R_{0}(\bm{x}_2, \bm{x}_2)}}\right),\label{eq:sbd}
    \vspace{-1mm}
\end{equation}
where $l \in\{1,2, \ldots, 2 m-1\}$ is the series phase shift index. $R_{k}(\cdot)$ is the cross-correlation function, as shown in ~\eqref{eq:cross_corelation}:
\begin{equation}
    R_{k}(\bm{x}_1, \bm{x}_2):=\left\{\begin{array}{ll}
\sum_{l=1}^{m-k} x_{1,l+k} \cdot x_{2,l}, & k \geq 0 \\
R_{-k}(\bm{x}_2, \bm{x}_1), & k<0
\end{array}\right.,\label{eq:cross_corelation}
\end{equation}
where $k$ is the relative phase shift index between $\bm{x}_1$, $\bm{x}_2$. From the cross-correlation property, $\phi_{\textit{shape}}\in[0,2]$, and the smaller $\phi_{\textit{shape}}$ indicates a higher shape similarity between $\bm{x}_1$ and $\bm{x}_2$. 

We posit that the similarity of the KPIs' intensity is predominantly reflected through the similarity of their peak intensity. As such, $\phi_{\textit{intensity}}(\cdot)$ computes the ratio of the peak mean values between $\bm{x}_1$,$\bm{x}_2$, which can be expressed as follows:
\vspace{-1mm}
\begin{equation}
    \phi_{\textit{intensity}}(\bm{x}_1,\bm{x}_2):=\exp\{{\frac{-1}{|q(\bm{x}_1,\bm{x}_2)-1|}}\},\label{eq:esbd_intensity}
    \vspace{-1mm}
\end{equation}
where $q(\bm{x}_1$, $\bm{x}_2)$ is the peak ratio function, described by the similarity between peaks of $\bm{x}_1$ and $\bm{x}_2$, as shown in:
\vspace{-1mm}
\begin{equation}
    q(\bm{x}_1,\bm{x}_2):=\max(\frac{\frac{1}{n_1}\sum v_{\bm{x}_1}^{i}}{\frac{1}{n_2}\sum v_{\bm{x}_2}^{i}}, \frac{\frac{1}{n_1}\sum v_{\bm{x}_2}^{i}}{\frac{1}{n_2}\sum v_{\bm{x}_1}^{i}})\label{eq:intensity_ratio},
    \vspace{-1mm}
\end{equation}
where $v_{\bm{x}_1}^{i}, v_{\bm{x}_2}^{i}$ are the peaks value of $\bm{x}_1,\bm{x}_2$, and $n_x, n_y$ are the peak number of $\bm{x}_1,\bm{x}_2$ respectively. In this context, a smaller $\phi_{\textit{intensity}}$ indicates a higher similarity between $\bm{x}_1,\bm{x}_2$ in terms of intensity. Additionally, $\phi_{\textit{intensity}}$ is more sensitized when $\bm{x}_1,\bm{x}_2$ exhibit similar intensity levels. Here, $q$ functions as a temperature parameter.

In summary, $\phi_{\textit{esbd}}$ lies within the range of $[0,2]$. Lower values of $\phi_{\textit{esbd}}$ suggest higher similarity between KPIs, modulated by the intensity factor $\alpha$. Generally, we set the intensity factor $\alpha$ to $0.5$ and use $1.0$ as the ESBD similarity threshold.

\subsection{pre-processing}\label{pre-processing}
The pre-processing $p(\cdot)$ aims to transform the raw workload KPIs $\tilde{\mathbf{X}}$ into incremental, low-noise, standardized KPIs $\mathbf{X}$. Toward this end, we sequentially apply five pre-processing steps: imputation $p_{\textit{im}}(\cdot)$, differentiation $p_{\textit{diff}}(\cdot)$, resampling $p_{\textit{rs}}(\cdot)$, denoising $p_{\textit{dn}}(\cdot)$ and standardisation $p_{\textit{std}}(\cdot)$.  It's important to note here that the KPIs that are utilized for the final evaluation do not undergo the standardization step $p_{\textit{std}}(\cdot)$.

\noindent\textbf{Imputation}. 
Given the typically low probability of KPI values missing, a straightforward and efficient method is used to address this issue. Specifically, cubic spline interpolation is applied to segments with missing values, as follows:
\begin{equation}
    \widetilde{\mathbf{X}}^{(\textit{im})}_{i, j:k} = p_{\textit{im}}(\widetilde{\mathbf{X}}_{i, j:k}, l):=\mathrm{CubicSpline}(\widetilde{\mathbf{X}}_{i, j-l:k+l}) \label{eq:imputation},
\end{equation}
where $\widetilde{\mathbf{X}}_{i, j:k}$ is $\widetilde{\mathbf{X}}_{i}$ with the missing value segment $(j:k)$, and the non-missing values at segment $(j-l:k+l)$ are the interpolation point, while $l$ is the interpolation interval length.

\noindent\textbf{Differentiation}. 
In order to effectively highlight the incremental features that emphasize real-time characteristics, we apply the first-order differential to cumulative KPIs, as follows:
\vspace{-1mm}
\begin{equation}
    \begin{aligned}
        \widetilde{\mathbf{X}}^{(\textit{diff})}_{i} = p _{\textit{diff}}(\widetilde{\mathbf{X}}^{(\textit{im})}_{i})
        := \widetilde{\mathbf{X}}^{(\textit{im})}_{i,1:m-1} - \widetilde{\mathbf{X}}^{(\textit{im})}_{i,0:m-2}\label{eq:diff}.
    \end{aligned}
    \vspace{-1mm}
\end{equation}

\noindent\textbf{Resampling}. 
For an analysis aligning with certain levels of granularity, specific periods—such as macroscopic time intervals—should be used for macroscopic periods. To synchronize these time intervals with the corresponding periods, resampling is performed on KPIs. Suppose $\tau$ signifies a suitable time interval and $\Delta t$ is the default time interval of the KPI. Now, the resampling process would proceed as follows:
\vspace{-1mm}
\begin{equation}
    \widetilde{\mathbf{X}}_i^{(\textit{rs})} = p_{\textit{rs}}(\widetilde{\mathbf{X}}_i^{(\textit{diff})}):={\Bigg|\Bigg|_{j=0}^{m'}}{\sum_{k=0}^{l}{\widetilde{\mathbf{X}}_{i,j+k}^{(\textit{diff})}}}
    \label{eq:resample},
    \vspace{-1mm}
\end{equation}
where $l=\lfloor\frac{\tau}{\Delta t}\rfloor$ is the ratio of resample, and $m' = \lfloor\frac{m}{l}\rfloor$ is the length of KPIs after resample operation while $\widetilde{\mathbf{X}}^{(\textit{rs})} \in\mathbb{R}^{n\times m'}$.

\noindent\textbf{Denoising}. 
To address the random noise inherent in the KPIs, we will apply the Kalman filter\cite{welch1995introduction} to the KPIs $\widetilde{\mathbf{X}}^{(\textit{rs})}$:
\vspace{-1mm}
\begin{equation}
    \widetilde{\mathbf{X}}^{(\textit{dn})}=p_{de}(\widetilde{\mathbf{X}}^{(\textit{rs})}):=\mathrm{KalmanFilter}(\widetilde{\mathbf{X}}^{(\textit{rs})})\label{eq:filter}.
    \vspace{-1mm}
\end{equation}

\noindent\textbf{Standardisation}. 
There exist potential relationships between KPIs at differing scales and KPI compression only considers similarities in shape, not differing magnitudes. Therefore, to align and compare these KPIs, standardisation is applied as:
\vspace{-1mm}
\begin{equation}
    \mathbf{X}_i = p_{\textit{std}}(\widetilde{\mathbf{X}}_i^{(\textit{dn})}) := \frac{\widetilde{\mathbf{X}}_i^{(\textit{dn})}-\mu_i}{\sigma_i}
    \label{eq:stand},
    \vspace{-1mm}
\end{equation}
where $\mu_i$ is the mean of $\widetilde{\mathbf{X}}_i^{(\textit{dn})}$ and $\sigma_i$ is the standard deviation.

\subsection{Compression}\label{Compression}
Microservice systems typically host an extensive array of KPIs, which can complicate individual analyses for evaluation purposes. Moreover, different workload simulation tasks usually contain varied business logic, enhancing the representativeness of business-related KPIs for these workload patterns. However, this does not imply the irrelevance of business-independent KPIs. To address this, we design a KPI filter and a KPI cluster mechanism to categorize all KPIs $\mathcal{K}$ into strongly correlated KPIs $\mathcal{K}_s$ and weakly correlated KPIs $\mathcal{K}_w$, based on their relevance to business operations. Specifically, the KPI filter leverages domain knowledge and chaos experiments\cite{basiri2016chaos} to derive $\mathcal{K}_s$, while the KPI clustering mechanism selects a subset of representative weakly correlated KPIs $\mathcal{K}_w$ from $\mathcal{K}$.

\subsubsection{Filter}\label{Filter} The KPIs filter is bifurcated into coarse filter by domain knowledge and refined filter by chaos experiments.
\noindent\textbf{Domain Knowledge Filter}. 
In microservice systems, KPIs are typically distinguished by fine-grained label attributes and coarse-grained domain attributes. The former encapsulates specific characteristics, while the latter reflects categorical traits, such as application source (like ENVOY or ISTIO), and monitoring level (like NODE or NETWORK). Accordingly, through the lens of domain attributes, we can examine the relevance of KPIs to the business, thereby establishing a coarse-grained filter, denoted as DKF. This filter yields the coarse-grained KPI set $\mathcal{K}_c$.

\noindent\textbf{Chaos Experiments Filter}. 
We deploy chaos experiments to extract the fine-grained KPI set $\mathcal{K}_s$ from $\mathcal{K}_c$. This is part of a refined filtering process, denoted as CEF, which injects the same workload under varying perturbations. We model these correlations by calculating the steady-state deviations of KPIs relative to business-independent faults. The divergence between these steady-states informs us about the relevance of KPIs, aiding in their fine-graining.

Our chaos experiment is organized into two control and one experimental group. The control groups provide a steady-state baseline for the system, maintained without the introduction of any perturbations. In contrast, the experimental group is subjected to a set of business-independent perturbations, denoted $\bm{\delta}$. For a specific KPI $k$, instances $\bm{x_1}$ and $\bm{x_2}$ stand for the control group's KPI measurements, while $\tilde{\bm{x}}$ denotes the one from the experimental group. The interrelation among $\bm{x_1}, \bm{x_2}, \tilde{\bm{x}}$ can be mathematically formulated as follows:
\vspace{-1mm}
\begin{equation}
    \begin{aligned}
        \bm{x_2}&=\bm{x_1} + \bm{\varepsilon},\\
        \tilde{\bm{x}}&=\bm{x_1}+\lambda\bm{\delta}  + \tilde{\bm{\varepsilon}},\\
        \bm{\varepsilon}&\neq \tilde{\bm{\varepsilon}},
    \end{aligned}
\vspace{-1mm}
\end{equation}
where $\lambda\in[0, 1]$ is the degree of the $\bm{\delta}$ perturbations, $\bm{\varepsilon}, \tilde{\bm{\varepsilon}}$ are the inherent system errors.The term $\lambda \bm{\delta}$ denotes the steady-state offset in response to perturbations; a larger value of $\lambda$ implies that $k$ is more susceptible to the perturbations $\bm{\delta}$ and holds a lesser relevance to the business operations. Despite this, precise calculations of $\lambda$ and $\bm{\delta}$ or deriving the exact value of $\lambda\bm{\delta}$, are challenging tasks. We therefore estimate $\lambda\bm{\delta}$ by quantifying the relevance to business operations through the contrastive analysis of ESBD among $\bm{x_1}, \bm{x_2}, \tilde{\bm{x}}$. 
Let $\phi_{\textit{esbd}}(\bm{x_1}, \bm{x_2}), \phi_{\textit{esbd}}(\bm{x_1}, \tilde{\bm{x}}), \phi_{\textit{esbd}}(\bm{x_2}, \tilde{\bm{x}})$ symbolize intra-group and inter-group ESBD respectively, denoted for simplicity as $\Phi_0, \Phi_1, \Phi_2$. We introduce a perturbation coefficient as a measure of the KPI's perturbability, expressed as:
\begin{equation}
    r(\Phi_0, \Phi_1, \Phi_2):=\max(\log{\frac{\exp^{\Phi_1+\Phi_2}}{\exp^{2\Phi_0}}}, \log{\frac{\exp^{2\Phi_0}}{\exp^{\Phi_1+\Phi_2}}}).
\end{equation}

When the perturbation coefficient $r$ is less than a specific threshold  $\gamma\in(0, +\infty)$, $k$ is considered a strong correlation KPI. We set $\gamma=1$ for the following experiments.

Iterate all the $k\in \mathcal{K}_c$, and individually verify their compliance with the aforementioned conditions. This process identifies the set of strongly correlated KPIs $\mathcal{K}_s$ and concurrently, determines the set of weakly correlated KPIs $\tilde{\mathcal{K}}_w$, which is the complement of $\mathcal{K}_s$ in $\mathcal{K}$.

\subsubsection{Cluster}\label{Cluster} To explore potential connections within KPIs, we perform clustering on the business-independent KPIs based on their shape, taking the centroid KPI in each cluster to form the weak correlation KPIs set ${\mathcal{K}}_w$.

\renewcommand{\algorithmicrequire}{\textbf{Input:}}
\renewcommand{\algorithmicensure}{\textbf{Output:}}
\begin{figure}[htbp]
    \vspace{-6mm}
    \begin{algorithm}[H]
        \scriptsize
        \caption{Heuristic Density Estimation}
        \label{ag:density_estimation}
        \begin{algorithmic}[1]
            \Require $\bm{k}_{\textit{dis}}$: DTW-based KNN curve; 
            \Statex$\textit{max\_raidus}$: upper bound of the density radius; 
            \Statex$\textit{left}$: smallest index satisfies $\textit{k\_dist}[\textit{left}]\leq\textit{max\_raidus}$;
            \Statex$\textit{right}$: largest index;
            \Statex$\textit{len\_thresh}$: threshold of segment length;
            \Statex$\textit{slope\_thresh}$: threshold of slope;
            \Statex$\textit{slope\_diff\_thresh}$: threshold of difference between slopes;
            \Ensure $\textit{radius\_index}$: index of largest candidate radius.
            \Function{RadiusEstimation}{$\bm{k}_{\textit{dis}}$, $\textit{left}$, $\textit{right}$}
            \If{$\textit{right} - \textit{left} < \textit{len\_thresh}$}
                \State \Return //Search area is too small;
            \EndIf
            \If{$\textit{left} < 0$ or $\textit{right} < 0$}
                \State \Return // No flat portion from the last step
            \EndIf
            
            \State $\textit{index}\gets -1$, $\textit{radius\_index}\gets -1$, $\textit{diff}\gets \textit{max\_raidus}$;
        
            \For{$i=(\textit{left}+1) \to i=\textit{right}$}
                \State $\textit{left\_slope}\gets -(\bm{k}_{\textit{dis}}[i] - \bm{k}_{\textit{dis}}[\textit{left}])/(i-\textit{left})$;
                \State $\textit{right\_slope}\gets -(\bm{k}_{\textit{dis}}[i] - \bm{k}_{\textit{dis}}[\textit{right}])/(i-\textit{right})$;
                \If{$\textit{left\_slope}$ $>$ $\textit{slope\_thresh}$ \textbf{and} $\textit{right\_slope}$ $>$ $\textit{slope\_thresh}$} 
                    \State \textbf{continue}; // Prune the steep portion  
                \EndIf
        
                \If{$|\textit{left\_slope}-\textit{right\_slope}|<\textit{diff}$}
                    \State $\textit{diff}\gets |\textit{left\_slope} - \textit{right\_slope}| $, $\textit{index}\gets i$;
                \EndIf
            \EndFor
        
            \If{$\textit{diff}<\textit{slope\_diff\_thresh}$}
                \State $\textit{radius\_index}\gets\textit{index}$
            \EndIf
            
            // Divide and Conquer 
            \State $\textit{radius\_index} \gets\min(\textit{radius\_index},$\Call{RadiusEstimation}{$\bm{k}_{\textit{dis}}$, $\textit{index}$, $\textit{right}$}$)$;    
            \For{$i=(\textit{index}) \to i=\textit{left}$}
                \If{$\bm{k}_{\textit{dis}}[i] - \bm{k}_{\textit{dis}}[\textit{index}]>\textit{diff}$}
                    \State $\textit{index}\gets i$;
                    \State \textbf{break};// Search forward for the next flat portion.
                \EndIf
            \EndFor
            \State $\textit{radius\_index} \gets\min(\textit{radius\_index},$\Call{RadiusEstimation}{$\bm{k}_{\textit{dis}}$, $\textit{left}$, $\textit{index}$}$)$;
        
            \Return $\textit{radius\_index}$
            \EndFunction
        \end{algorithmic}
    \end{algorithm}
\vspace{-6mm}
\end{figure}

Inspired by the prior KPI clustering algorithm\cite{li2018robust, wang2021rapid, yu2019unsupervised}, we design a heuristic approach for density estimation to a DTW-based DBSCAN clustering model, providing adaptive density radius, as shown in Algorithm~\ref{ag:density_estimation}.
The crux of our heuristic density estimation is the identification of feasible density radius candidates on flat portions of the K-Nearest-Neighbor (KNN) curve, which represents the descending order DTW distance between each KPI instance and its K-Nearest-Neighbor, denoted as $\bm{k}_{\textit{dis}}$.
Specifically, for any given point on the curve, we determine its flat portion status – indicative of a potential radius – by comparing slope differences at the left and right endpoints. We utilise a divide-and-conquer strategy to iterate over both sides of the point, identifying all potential candidates and selecting the largest as the estimated radius. This process is optimised through pruning for recursion reduction.
Additionally, we adopt the upper boundary pruning methodology\cite{silva2016speeding}, alongside the search constraints recommended\cite{sakoe1978dynamic}, to expedite DTW computation. We further enhance DBSCAN clustering efficiency by chunking KPIs within their domain attributes.

Upon obtaining a series of clusters through the DTW-based DBSCAN, we compute the centroid of each cluster as a representative KPI for evaluation, expressed as follows:
\vspace{-1mm}
\begin{equation}
    \bm{z}_{i}=\arg\min_{\bm{x}\in  C_i}\sum_{\bm{y}\in C_i}\text{DTW}(\bm{x},\bm{y}),
    \vspace{-1mm}
\end{equation}
where $C_i$ is one of the clusters from DBSCAN, then, the KPIs $k_i\in\tilde{\mathcal{K}}_w$ represented by the centroid of each cluster form a reduced weak correlation KPI set $\mathcal{K}_w$.
\subsection{Evaluation}\label{evaluation}
Leveraging the strong and weak correlation KPI sets $\mathcal{K}_s$ and $\mathcal{K}_w$, which we obtain from above steps respectively, we apply ESBD to calculate the similarity between the original and simulated workload by aggregation with the statistical importance weights for evaluation.

\noindent\textbf{Weak Correlation Evaluation}. Given the $i^{\text{th}}$ KPI in $\mathcal{K}_w$, whose instances under original and simulated workload are $\bm{x}_{\mathcal{K}_w^{(i)}},\bm{x}_{\mathcal{K}_w^{(i)}}^{'}$, we calculates the ESBD between the two, i.e., $\phi_{\textit{esbd}}(\bm{x}_{\mathcal{K}_w^{(i)}},\bm{x}_{\mathcal{K}_w^{(i)}}^{'})$, denoted as $\Phi_{\mathcal{K}_w^{(i)}}$, and then aggregated as:
\vspace{-1mm}
\begin{equation}
    \Phi_{\mathcal{K}_w}:=\frac{1}{|\mathcal{K}_w|} \sum_{i=0}^{|\mathcal{K}_w|-1} \Phi_{\mathcal{K}_w^{(i)}}\label{eq:weak_evaluation},
    \vspace{-1mm}
\end{equation}
in which the arithmetic mean of $\Phi_w$ characterises the distance between the original and the simulated workload, then normalized as a weak correlation evaluation score: 
\vspace{-1mm}
\begin{equation}
    E_{\mathcal{K}_w}=\frac{\mu_{\mathcal{K}_w}}{\mu_{\mathcal{K}_w}+\Phi_{\mathcal{K}_w}},
    \vspace{-1mm}
\end{equation}
where $\mu_{\mathcal{K}_w} \in (0, 2)$ is the similarity threshold of the weak correlation evaluation.

\noindent\textbf{Strong Correlation Evaluation}. 
As strong correlation KPIs monitor the state of each individual service, variances in these metrics mirror differences in workload patterns. To evaluate these KPIs, which form the foundation for workload simulation evaluation, we need a finer-grained aggregation process. Workloads generally include calls to a multiplicity of specific services. Thus, based on the types of services, we partition $\mathcal{K}_s$ into several services subsets corresponding to individual services, denoted by $\mathcal{K}_s^{(i)}$, such that $\bigcup _{i=1}^{n_t}\mathcal{K}_{s}^{(i)}=\mathcal{K}_s$, where $n_t$ is the number of service type. We also employ the frequency of service calls as the importance weight $\omega_i$, satisfying $\sum_{i=1}^{n_t}\omega_i=1.0$. Given the strong correlation KPIs $k_j\in \mathcal{K}_s$, we aggregate the ESBD for evaluation as follows:
\vspace{-1mm}
\begin{equation}
    \Phi_{\mathcal{K}_s}:=\sum_{i=1}^{n_t} \omega_{i}\left(\frac{1}{n_t^{(i)}} \sum_{j=1}^{n_t^{(i)}} \Phi_{j}\right), n_t^{(i)} = |\mathcal{K}_s^{(i)}|.
    \vspace{-1.5mm}
\end{equation}
Then, normalized as follows:
\vspace{-1mm}
\begin{equation}
    E_{\mathcal{K}_s}=\frac{\mu_{\mathcal{K}_s}}{\mu_{\mathcal{K}_s}+\Phi_{\mathcal{K}_s}},
    \vspace{-1.5mm}
\end{equation}
in which $\mu_{\mathcal{K}_s}$ is the similarity threshold of the strong correlation evaluation. In general, we set $\mu_{\mathcal{K}_w} = \mu_{\mathcal{K}_s}$.

Finally,  $E_{\mathcal{K}_w}$ and $E_{\mathcal{K}_s}$ were weighted by the correlation factor $\beta\in[0,1]$ to obtain a single similarity evaluation $E$ of the quality of the workload simulation, expressed as:
\vspace{-1mm}
\begin{equation}
    E=h_{\beta}(\mathcal{K}_w, \mathcal{K}_s):=(1-\beta) E_{\mathcal{K}_w}+\beta E_{\mathcal{K}_s}.
\end{equation}

\section{EXPERIMENTS}
In this section, we conduct empirical experiments to demonstrate the effectiveness of ESBD and KEWS. We aim to address the following three research questions: \textbf{RQ1:} How effective is ESBD for describing similarities in shape and strength? \textbf{RQ2:} How effective is KEWS for compressing KPIs with filter and cluster? \textbf{RQ3:} How accurate is KEWS for evaluating the quality of workload simulation?

\subsection{Experimental Setup}
\noindent\textbf{Benchmark System}. Our experiments are conducted in an open-source microservices application, \textit{Hipstershop}\footnote{https://github.com/GoogleCloudPlatform/microservices-demo}, which is a typical online e-commerce application widely used in the research of microservice operations. 

\begin{figure}[htbp]
    \centering
    \includegraphics[width=\columnwidth]{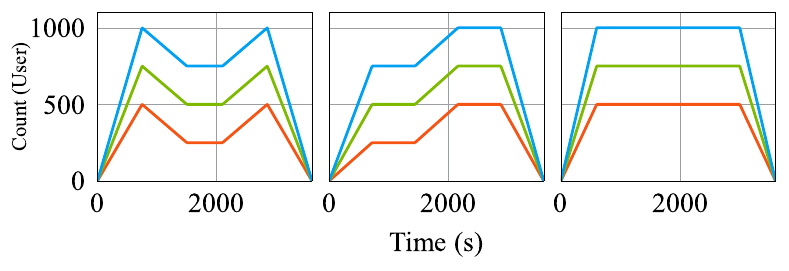}
    \vspace{-6mm}
    \caption{Workload curve setting}
    \label{fig:workload_setting}
\end{figure}

\noindent\textbf{Workload Simulation}.
We use \textit{K6} to generate real-world workloads on Hipstershop as the original workload $\mathcal{W}$. As shown in Fig.~\ref{fig:workload_setting}, we set up a total of nine workload scenarios with three shapes ($\mathcal{S}_a, \mathcal{S}_b, \mathcal{S}_c$) and three intensities ($\mathcal{I}_1, \mathcal{I}_2, \mathcal{I}_3$). We employ LWS\cite{han2023lws} to generate the simulated workload $\mathcal{W}^{'}$ and capture original log data for LWS by \textit{Elasticsearch}\footnote{https://www.elastic.co/cn/elasticsearch/} and \textit{Filebeat}\footnote{https://www.elastic.co/cn/beats/filebeat}, then capture KPIs data by \textit{Prometheus}\footnote{https://prometheus.io/docs/introduction/overview/}.

\noindent\textbf{Baselines}. For similarity metric, we compare ESBD with DTW\cite{berndt1994using} and SBD\cite{paparrizos2015k}. For workload simulation evaluation, we compare KEWS with WESSBAS\cite{vogele2018wessbas}.


\subsection{Effectiveness of ESBD (RQ1)}
\begin{figure}[htbp]
    \vspace{-3mm}
    \begin{subfigure}{0.49\columnwidth}
        \centering
        \includegraphics[width=1.0\linewidth]{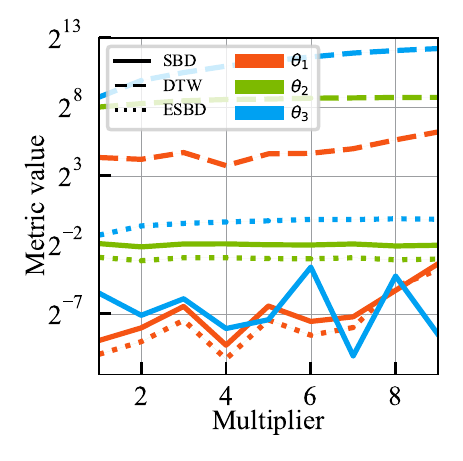}
        \vspace{-8mm}
        \caption{$\log_2$ scale}
        \label{fig:metric_result_1}
    \end{subfigure}
    \begin{subfigure}{0.49\columnwidth}
        \centering
        \includegraphics[width=1.0\linewidth]{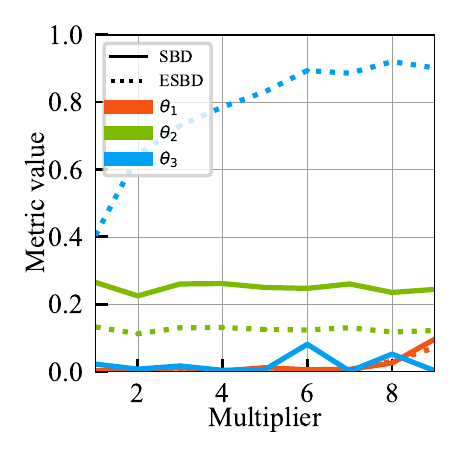}
        \vspace{-8mm}
        \caption{linear scale}
        \label{fig:metric_result_2}
    \end{subfigure}
    \caption{Trend of metric with change of $\theta_1$, $\theta_2$, $\theta_3$ }
    \label{fig:tsgen_result}
    \vspace{-5mm}
\end{figure}

Since the open-source KPIs dataset makes it difficult to meet the demand for shape and intensity control, we utilize the KPIs generation algorithm TSAGEN\cite{wang2021tsagen} to conduct the validation experiments to compare the trend between SBD and DTW. Fig.~\ref{fig:tsgen_result} reports the trend of our method ESBD and other baselines with shape change and intensity change, where $\theta_1, \theta_2$ are the shape factors and $\theta_3$ is the intensity factor and the horizontal axis is the multiplier of the factors.

As can be observed, the three similarity metrics have similar trends in the $log_2$ scale, as they can both capture shape features, while DTW has much larger values. In linear scale, Fig.~\ref{fig:metric_result_2} shows ESBD's ability to capture intensity features and the same to SBD in terms of shape, which validates the effectiveness of ESBD, compared to DTW and SBD.

\subsection{Effectiveness of Compression (RQ2)}
\begin{table}[htbp]
    \vspace{-3mm}
    \caption{Strong correlation domain attributes of KPIs}
    \label{tab:domain}
    \footnotesize
    \centering
    \setlength{\tabcolsep}{3mm}{
        \begin{tabular}{c|c}
            \hline
            \textbf{Domain} & \textbf{Description} \\
            \hline
            container & Kubernetes Resources \\
            \hline
            grpc & Kubernetes application, monitoring grpc service \\
            \hline
            http & Kubernetes application, monitoring http requests \\
            \hline
            istio & Kubernetes component, for workload management\\
            \hline
            node & Kubernetes Resources\\
            \hline
        \end{tabular}
    }
    \vspace{-3mm}
\end{table}
Based on the principle of locality and the domain knowledge of \textit{Kubernetes}, we summarise the strong correlation KPIs as shown in Table.~\ref{tab:domain}, including \textit{Kubernetes} resources and application. Then, 
We employ an open-source tool \textit{ChaosMesh} to conduct our chaos experiment. As shown in Table.~\ref{tab:chaos}, \textbf{grpc} and \textbf{istio} are the strong correlation domain KPIs described by $\gamma = 0.1$ and $r$, which are much smaller than others. 
\newcommand\Tiny{\fontsize{6}{6}\selectfont}
\begin{table}[htbp]
    \caption{The perturbation coefficient $r$ of chaos experiment}
    \label{tab:chaos}
    \scriptsize
    \centering
    \begin{tabularx}{\linewidth}{c|XXX|XXX|XXX|c}
        \hline
        \multirow{2}{*}{\textbf{Domain}} & \multicolumn{3}{c|}{$\mathcal{I}_1$} & \multicolumn{3}{c|}{$\mathcal{I}_2$} & \multicolumn{3}{c|}{$\mathcal{I}_3$} & \multirow{2}{*}{Avg.} \\ \cline{2-10}
                                & $\mathcal{S}_a$          & $\mathcal{S}_b$         & $\mathcal{S}_c$         & $\mathcal{S}_a$          & $\mathcal{S}_b$         & $\mathcal{S}_c$         & $\mathcal{S}_a$          & $\mathcal{S}_b$         & $\mathcal{S}_c$         &                          \\ \hline
         container               & \Tiny1.202	 &\Tiny1.176	&\Tiny1.186	&\Tiny1.170	&\Tiny1.173	&\Tiny1.154	&\Tiny1.167	&\Tiny1.149	&\Tiny1.168	&1.172 \\ \hline
        \textbf{grpc}	&\Tiny1.002	&\Tiny1.002	&\Tiny1.003	&\Tiny1.003	&\Tiny1.004	&\Tiny1.005	&\Tiny1.002	&\Tiny1.002	&\Tiny1.001	&\textbf{1.003}\\ \hline
        http	&\Tiny2.337	&\Tiny2.012	&\Tiny1.795	&\Tiny3.719	&\Tiny4.335	&\Tiny1.354	&\Tiny4.985	&\Tiny1.000	&\Tiny1.000	&2.504\\ \hline
        \textbf{istio}	&\Tiny1.064	&\Tiny1.133	&\Tiny1.090	&\Tiny1.078	&\Tiny1.088	&\Tiny1.084	&\Tiny1.081	&\Tiny1.123	&\Tiny1.086	&\textbf{1.092}\\ \hline
        node	&\Tiny1.189	&\Tiny1.183	&\Tiny1.182	&\Tiny1.166	&\Tiny1.177	&\Tiny1.164	&\Tiny1.177	&\Tiny1.174	&\Tiny1.177	&1.177\\ \hline
    \end{tabularx}
    \vspace{-3mm}
\end{table}

By analysing the KPI of each domain attribute one by one manually, we can confirm that the \textbf{grpc} and \textbf{istio} are strong correlation domain KPIs recording calls of services and GRPC requests, while container, node, and HTTP are the weak correlation that records system resource status, and show the effectiveness of the filter.

\begin{table}[htbp]
    \caption{Silhouette coefficient of cluster}
    \vspace{-3mm}
    \label{tab:cluster_result}
    \footnotesize
    \begin{center}
        \begin{tabular}{c|cccc}
        \hline
        \multirow{2}{*}{$\textbf{n\_sample}$} & \multicolumn{4}{c}{$\textbf{max\_radius}$} \\ 
                        \cline{2-5} 
                         & $\textbf{0.5}$    & $\textbf{1.0} $   & $\textbf{1.5}$    & $\textbf{2.0} $   \\ 
                         \hline
                        250	 &0.937	    &0.873	   &0.805	  &0.733 \\ 
                        \hline
                        500	 &0.899	    &0.872	   &0.785	  &0.813 \\ 
                        \hline
                        750	 &0.864	    &0.817	   &0.783	  &0.824 \\ 
                        \hline
                        1000	&0.867	&0.813	   &0.791	&0.827 \\ 
                        \hline
        \end{tabular}
    \end{center}
    \vspace{-5mm}
\end{table}

Then, we apply the down-sampling method under the domain to accelerate the cluster, and after calculating the centroid of each cluster, the other KPIs are assigned to the clusters with the smallest DTW from the centroid. Table.~\ref{tab:cluster_result} reports the silhouette coefficient result on different sample numbers and \textit{max\_radius}, which shows promising results in various settings. 


\subsection{Effectiveness of Evaluator (RQ3)}
Based on the data $\mathcal{D}$ and $\mathcal{D}^{'}$ obtained by injecting the original workload and simulated workload, we additionally use the data from two same original workloads $\mathcal{S}_{a}$, namely the zero shape $\mathcal{S}_{o}$ for more intuitive evaluation, where the evaluation of $\mathcal{S}_{o}$ should be of high similarity. Table.~\ref{tab:eva_result} reports the result of evaluation between KEWS and WESSBAS by averaging from $\mathcal{I}_1$ to $\mathcal{I}_3$. Compared to WESSBAS, KEWS achieves better performance on $\mathcal{S}_{o}, \mathcal{S}_{a}, \mathcal{S}_{b}$, validating the effectiveness of KEWS. We make other observations as follows. Firstly, the decrease of the KEWS similarity with decreasing shape complexity indicates ESBD is more sensitive to error in smooth curves. Secondly, the small fluctuations in the similarity of WESSBAS reflect its shortcomings in evaluating only static statistics, and its inability to perceive differences at the system level. Thirdly, the overall low value of $\mathcal{K}_w$ also confirms the effective partition of $\mathcal{K}_w$ and $\mathcal{K}_s$.
\begin{table}[htbp]
    \vspace{-2mm}
    \caption{KPIs evaluation result($\mu_{\mathcal{K}_s}=\mu_{\mathcal{K}_w}=0.2, \beta=0.9$)}
    \label{tab:eva_result}
    \footnotesize
    \centering
    \begin{tabular}{c|cccc}
        \hline
        \multirow{2}{*}{\textbf{Method}}  &  \multicolumn{4}{c}{\textbf{Workload Type}}\\
        \cline{2-5}
         & $\mathcal{S}_o$ & $\mathcal{S}_a$ & $\mathcal{S}_b$ & $\mathcal{S}_c$ \\
        \hline
        WESSBAS  & 95.3 & 94.3 & 94.8 & 94.5 \\ 
        \hline
        KEWS($E_{\mathcal{K}_w}$) & 93.7 & 95.5  & 94.5  & 91.9 \\
        \hline
        KEWS($E_{\mathcal{K}_s}$) & 99.1 & 96.3  & 95.3  & 93.3 \\
        \hline
        KEWS($E$) & 98.5  & 96.1  & 95.2  & 93.0 \\
        \hline
    \end{tabular}
    \vspace{-3mm}
\end{table}
\section{Conclusion}
In this paper, we propose ESBD and KEWS to evaluate the similarity between workloads, aiming at the challenges of large-scale, complex characteristics of KPIs in the production environment. Our ESBD measures the similarity of KPIs in terms of shape and intensity and our KEWS, composed of three modules—preprocessing, compression, and evaluation—differentiates between sets of strong and weak correlation KPIs and then evaluates the simulated workload employing ESBD. Extensive experiments demonstrate the superiority of ESBD and KEWS.

\bibliographystyle{IEEEtran}
\bibliography{ref}

\end{document}